\documentclass[11pt,letterpaper]{article}
\usepackage[dvips,linkcolor=blue,urlcolor=blue,citecolor=blue,linktocpage,bookmarks,pdfstartview=FitBH, breaklinks = true]{hyperref}
\usepackage{fancyhdr}
\usepackage[all]{hypcap}
\usepackage{latexsym}
\usepackage{verbatim}
\usepackage{setspace}
\usepackage[centertags]{amsmath}
\usepackage{amsfonts}
\usepackage{amssymb}
\usepackage{amsthm}
\usepackage{mathrsfs}
\usepackage{newlfont}
\usepackage{natbib}
\usepackage{graphicx}
\usepackage{lscape}
\usepackage{bm}
\usepackage{appendix}
\usepackage{rotating}
\usepackage{url}
\usepackage{float}
\usepackage{enumerate}

\numberwithin{equation}{section}

\theoremstyle{plain}
\newtheorem{theorem}{Theorem}[section]

\theoremstyle{definition}

\theoremstyle{remark}

\numberwithin{equation}{section}

\newcommand{\ra}{\rightarrow}

\newcommand{\bbeta}{\bm{\beta}}
\newcommand{\hbbeta}{\hat{\bm{\beta}}}

\def\X{\bm{X}}
\def\Y{\bm{Y}}
\def\x{\bm{x}}
\def\b0{\bm{0}}
\def\H{\bm{H}}
\def\h{\bm{h}}
\def\C{\bm{C}}

\def\b{\bm{b}}
\def\hbbetaUR{\bm{\hat\beta}^{\textrm{UR}}}
\def\hbbetaR{\bm{\hat\beta}^{\textrm{R}}}
\def\hbbetaSh{\bm{\hat\beta}^{\textrm{S}}}
\def\hbbetaS+{\bm{\hat\beta}^{\textrm{S+}}}

\def\D2{\Delta^2} 

\def\bvep{\bm{\varepsilon}}

\begin{document}

\singlespacing

\begin{center}
{\huge\bf Positive-shrinkage and Pretest Estimation in Multiple
Regression: A Monte Carlo study with Applications}
\end{center}

\vspace{2mm}

\centerline {SM Enayetur Raheem\footnote{Author for correspondence. Email: raheem@gmail.com} and S. Ejaz Ahmed}

\centerline{\small\it University of Windsor, Windsor, ON, Canada}

\medskip

\centerline {\today}
\vspace {4mm}

\centerline{\bf Abstract}

Consider a problem of predicting a response variable using a set of covariates in a linear regression model. If it is \emph{a priori} known or suspected that a subset of the covariates do not significantly contribute to the overall fit of the model, a restricted model that excludes these covariates, may be sufficient. If, on the other hand, the subset provides useful information, shrinkage method combines restricted and unrestricted estimators to obtain the parameter estimates. Such an estimator outperforms the classical maximum likelihood estimators. Any \emph{prior} information may be validated through preliminary test (or pretest), and depending on the validity, may be incorporated in the model as a parametric restriction. Thus, pretest estimator chooses between the restricted and unrestricted estimators depending on the outcome of the preliminary test. Examples using three real life data sets are provided to illustrate the application of shrinkage and pretest estimation. Performance of positive-shrinkage and pretest estimators are compared with unrestricted estimator under varying degree of uncertainty of the prior information. Monte Carlo study reconfirms the asymptotic properties of the estimators available in the literature.



\section{Introduction}\label{sec:intro}

Regression analysis is one of the most mature and widely applied
branch in statistics. Least squares estimation and related procedures,
mostly having a parametric flavor, have received considerable
attention from theoretical as well as application perspectives. Statistical models, both linear and non-linear, are used to obtain information about unknown parameters. Whether such model fits the data well or whether the estimated parameters are of much use depends on the validity of certain assumptions. In this setup, the estimates are obtained to have insights about the parameters. However, in many practical situations, it is the researchers who provide the estimation of the parameters utilizing the information contained in the sample and other relevant information. The ``other'' information may be considered as \emph{non-sample information} (NSI). This is also known as \emph{uncertain prior information} (UPI), or simply \emph{prior} information. The non-sample information may or may not positively contribute in the estimation procedure. Nevertheless, it may be advantageous to use the NSI in the estimation process when sample-information may be rather limited.

The quality of the fit and of the estimated parameters depend largely on the quality of the data used to obtain them. Only reliable information leads to useful results. However, in many practical situations, uncertainty arises as to whether the available information is of much use. It is widely accepted that in applied science, an experiment is often performed with some prior knowledge of the outcomes, or to confirm a hypothetical result, or to re-establish existing results.

With this keeping in mind, it is however, important to note that the consequences of incorporating non-sample information depend on the quality or usefulness of the information being added in the estimation process. Any uncertain prior information may be tested before they are incorporated in the model. Based on the idea of \citet{bancroft:1944}, uncertain prior information may be validated through preliminary test, and depending on the validity, may be incorporated in as a parametric restriction, and choose between the restricted or unrestricted estimation procedure depending on the outcome of the preliminary test.

Later, \citet{stein:1956} introduced shrinkage estimation. In this framework, the shrinkage estimator or Stein-type estimator takes a hybrid approach by shrinking the base estimator to a plausible alternative estimator utilizing the non-sample information if it proves to be useful.

\subsection{Review of Literature}

Since the beginning, shrinkage estimation have received considerable attention from the researchers. Since 1987, Ahmed and his co-researchers are among others who have analytically demonstrated that shrinkage estimators outshine the classical maximum likelihood estimator. Asymptotic properties of shrinkage and preliminary test estimators using quadratic loss function have been studied, and their dominance over the usual maximum likelihood estimators demonstrated in numerous studies in the literature. \cite{ahmed:1997} gave a detailed description of shrinkage estimation, and discussed large sample estimation techniques in a regression model with non-normal errors.

\cite{khan:ahmed:2003} considered the problem of estimating the coefficient vector of a classical regression model, and demonstrated analytically and numerically that the positive-part of Stein-type estimator, and the improved preliminary test estimator dominate the usual Stein-type, and pretest estimators, respectively.

Estimation of the mean vector of a multivariate normal distribution, under the uncertain prior information that component means are equal but unknown, was studied by \cite{khan:ahmed:2006}. \cite{ahmed:nicol:2010} among others, considered various large sample estimation techniques in a nonlinear regression model. Nonparametric estimation of the location parameter vector when uncertain prior information  about the regression parameters is available was considered by \cite{ahmed:saleh:1999:11:1}.

In this paper, we review positive shrinkage, and pretest estimators to compare their performance when certain information about a subset of the covariates are available \emph{a priori}. In particular, we apply shrinkage estimation on three real life data sets to show the usability of positive-shrinkage and pretest estimators for practical purposes.

\section{Statement of the Problem}\label{sec:statement:shrinkage-lasso}

Consider a regression model of the form
\begin{equation}\label{eq:full:model}
\Y=\X \bbeta+ \bvep,
\end{equation}
where $\Y=(y_1, y_2, \dots, y_n)'$ is a vector of responses, $\X$ is an $n \times p$ fixed design matrix, $\bbeta=(\beta_1,\dots, \beta_p)'$ is an unknown parameter vector and $\bvep = (\varepsilon_1, \varepsilon_2, \dots, \varepsilon_n)'$ is the vector of unobservable random errors, and the superscript ($'$) denotes the transpose of a vector or matrix.

We do not make any distributional assumption for the errors, only that $\bvep$s have a cumulative distribution function $F(\varepsilon)$ with $E(\bvep)=\b0$, and
$E(\bvep \bvep') = \sigma^2\bm{I}$, where $\sigma^2$  is finite. We make the following two assumptions, also called the regularity conditions
\begin{enumerate}[i)]
 \item $\displaystyle\mathop{\textrm{max}}_{1 \le i \le n} \x_i'(\X'\X)^{-1}\x_i \longrightarrow 0$ as $n \longrightarrow \infty$, where $\x_i'$ is the $i$th row of $\X$ \label{ass:assumption:1}
\item $\displaystyle\lim_{n \rightarrow \infty} \left(\frac{\X'\X}{n}\right) = \bm{C}_n$, where $\bm{C}_n$ is a finite positive-definite matrix.\label{ass:assumption:2}
\end{enumerate}

In our case, suppose that $\bm{\beta}$ may be partitioned as $\bm{\beta}=(\bm{\beta}'_1,\bm{\beta}'_2)'$. The sub-vectors $\bm{\beta}_1$ and $\bm{\beta}_2$ are assumed to have dimensions $p_1$ and $p_2$ respectively, and $p_1+p_2=p$, $p_i\ge 0$ for $i=1, 2$. Here, $\bm{\beta_1}$ is the coefficient vector for main effects, and $\bm{\beta_2}$ is a vector for ``nuisance'' effects.  We are essentially interested in the estimation of $\bm{\beta}_1$  when it is plausible that $\bm{\beta}_2$ do not contribute significantly in predicting the response. Such a situation may arise when there is over-modeling and one wishes to cut down the irrelevant part from the model (\ref{eq:full:model}). For example, in studying the relationship between the level of prostate specific antigen (PSA) and some clinical measures, the log cancer volume and log prostate weight can be considered as the main effects while age, log of benign  prostate hyperplasia amount, seminal vesicle invasion and others can be regarded as nuisance variables. In this situation, inference about $\bm{\beta_1}$ may benefit from shrinking the regression coefficients of the full model towards the restricted space while utilizing the available information contained in the nuisance covariates.
Thus, the parameter space can be partitioned, and it is plausible that $\bm{\beta}_2$ is near  some specified $\bm{\beta}^o_{2}$, which, without loss of generality, may be set to a null vector. The prior information about the subset of $\bbeta$ can be written in terms of a restriction, $\H\bbeta = \h$. Here, $\H$ is a known $p_2 \times p$ matrix and $\h$ is $p_2 \times 1$ vector of known constants.

%
%

\subsection{Organization of the Paper}
The paper is organized as follows. The statistical model is introduced in section 3. Shrinkage, positive-shrinkage, and pretest estimators are defined in this section. Examples using three real life data sets are presented in section 4. Positive-shrinkage and pretest estimators are obtained, and their performance are compared using cross-validation. Monte Carlo simulation study is described in section 5. Asymptotic bias and risk expressions for the shrinkage estimators are presented in section 6. Finally, conclusions and future directions are presented in section 7.

\section{The Model and Estimation Strategies}
The least-squares estimator of $\bbeta$ is given by
\[
 \hbbetaUR = (\X'\X)^{-1}\X'\Y = \C^{-1}\X'\Y,
\]
where $\C=(\X'\X).$ Under the restriction $\H\bbeta= \h$, the restricted estimator is given by
\[
 \hbbetaR = \hbbetaUR - \C^{-1}\H'(\H\C^{-1}\H')^{-1}(\H\hbbetaUR- \h),
\]
which is a linear function of the unrestricted estimator. Let us define the estimator of $\sigma^2$ by
\[
 s_e^2 = \frac{(\Y-\X\hbbetaUR)'(\Y-\X\hbbetaUR)}{n-p}.
\]
We may consider testing the restriction in the form of testing the null hypothesis
\[
 H_0: \H\bbeta = \h.
\]
The test statistic is defined by
\begin{equation}\label{eq:test:statistic}
 \psi_n = \frac{(\H\hbbetaUR - \h)'(\H\C^{-1}\H')^{-1}(\H\hbbetaUR-\h)}{s_e^2},
\end{equation}
which, under $H_0$, follows a chi-square distribution with $p_2$ degrees of freedom.

\subsection{Shrinkage Estimator}
\label{sec:estimation:strategies}

A Stein-type estimator (STE) $\hat{\bm{\beta}}^{\textrm S}_1$ of $\bm{\beta}_1$
can be defined as
\[
 \hbbeta_1^{\textrm{S}}= \hbbeta_1^{\textrm{R}} + (\hbbeta_1^{\textrm{UR}} - \hbbeta_1^{\textrm{R}}) \left\{1- \kappa\psi_n^{-1}\right\}, \mbox{ where }\kappa=p_2-2, \quad p_2 \ge 3.
\]
where $\psi_n$ is defined in (\ref{eq:test:statistic}).

One problem with STE is that its components may have a different sign
from the coordinates of $\hbbeta_1^{\textrm{UR}}$. This could happen if
$(p_2-1)\psi_n^{-1}$ is larger than unity. One possibility is when
$p_2=2$ and $\psi_n < 1.$ From the practical point of view, the change
of sign would affect its interpretability. However, this behavior does
not adversely affect the risk performance of STE. To overcome the
sign problem, we define a positive-rule Stein-type semiparametric
estimator (PSTE) by retaining the positive-part of the STE. A PSTE
has the form
\[
 \hbbeta_1^{\textrm{S+}}= \hbbeta_1^{\textrm{R}} +
(\hbbeta_1^{\textrm{UR}} - \hbbeta_1^{\textrm{R}}) \left\{1-
\kappa\psi_n^{-1}\right\}^{+}, \quad p_2 \ge 3
\]
where $z^{+}=\max(0,z)$.
Alternatively, this can be written as
\[
 \hbbeta_1^{\textrm{S+}}= \hbbeta_1^{\textrm{R}} +
(\hbbeta_1^{\textrm{UR}} - \hbbeta_1^{\textrm{R}}) \left\{1-
\kappa\psi_n^{-1}\right\}I(\psi_n< \kappa), \quad p_2 \ge 3.
\]
\cite{ahmed:2001} and others studied the
asymptotic properties of Stein-type estimators in various contexts.

\subsection{Preliminary Test Estimator}
The preliminary test estimator or pretest estimator for the regression parameter $\bbeta_1$ is obtained as
\begin{equation}
 \hbbeta^{\textrm{PT}}_1 = \hbbeta^{\textrm{UR}}_1 - (\hbbeta^{\textrm{UR}}_1 - \hbbeta^{\textrm{R}}_1)I(\psi_n < c_{n, \alpha}),
\end{equation}
where $I(\cdot)$ is an indicator function, and $c_{n, \alpha}$  is the upper $100(1-\alpha)$  percentage point of the test statistic $\psi_n$.

In a pretest estimation problem, the \emph{prior} information is tested before choosing the estimator for practical purposes, while shrinkage and positive-shrinkage estimator incorporates in the estimation process whatever \emph{prior} information is available.

Pretest estimator either accepts of rejects the restricted estimator ($\hbbeta^{\textrm{R}}_1$) based on whether $\psi_n < c_{n, \alpha}$, while shrinkage estimator is a smoothed version of the pretest estimator.

\section{Examples} \label{chap:sec:motivating:example}


In the following, we study three real life examples. For each data set, we fit linear regression models to predict the variable of interest form the available regressors. Shrinkage and pretest estimates are then obtained for the regression parameters. Performance of shrinkage and pretest estimators are assessed as per the criteria outlined in the following section.

\subsection{Assessment Criteria}

In shrinkage and pretest estimation, we utilize the full-model and  sub-model estimates, and combine them in a way that shrinks the least-squares estimates towards the sub-model estimates. In this framework, we utilize, if available, the information contained in the restricted subspace if they contribute significantly in predicting the response. However, in the absence of prior information about the nuisance subset, one might do usual variable selection to filter the nuisance subset out of the covariates. In that, one initiates the process with the model having all the covariates. Then the best subset may be selected based on AIC, BIC or other model selection criteria. Separate estimates from full- and restricted models are then combined to obtain shrinkage estimates. Finally, a model with shrunken coefficients is obtained, which reduces overall prediction error.

We obtain pretest and positive-shrinkage estimates using different sub-models. Performance of each pair of full- and sub-models was evaluated by estimating the prediction error based on $K$-fold cross validation. In a cross validation, the data set is randomly divided into $K$ subsets of roughly equal size. One subset is left aside, and termed as test data, while the remaining $K-1$ subsets, called training set, are used to fit the model. The fitted model is then used to predict the responses of the test data set. Finally, prediction errors are obtained by taking the squared deviation of the observed and predicted values in the test set.

We consider $K=5, 10$. Both raw cross validation estimate (CVE), and bias corrected cross validation estimate of prediction errors are obtained for each configuration.  The bias corrected cross validation estimate is the adjusted cross-validation estimate designed to compensate for the bias introduced by not using leave-one-out cross-validation \citep{tibshirani:tibshirani:2009}.

Since cross validation is a random process, the estimated prediction error varies across runs, and for different values of $K$. To account for the random variation, we repeat the cross validation process 5000 times, and estimate the average prediction errors along with their standard errors. The number of repetitions was initially varied, and settled with this as no noticeable variations in the standard errors were observed for higher values.

\subsection{Prostate Data}

\cite{hastie:tibshirani:friedman:2009} demonstrated various model selection techniques by fitting linear regression model to the prostate data. Specifically, the log of prostate-specific antigen ({\tt lpsa}) was modeled by the
log cancer volume ({\tt lcavol}), log prostate weight ({\tt lweight}), age ({\tt age}), log benign prostatic hyperplasia amount ({\tt lbph}), seminal vesicle invasion ({\tt svi}), log capsular penetration ({\tt lcp}), Gleason score ({\tt gleason}), and percentage Gleason scores 4 or 5 ({\tt pgg45}). The idea is to predict {\tt lpsa} from the measured variables.

The predictors were first standardized to have zero mean and unit standard deviation before fitting the model. Several model selection criteria and shrinkage methods were tried--details of which may be found in \citet[Table~3.3, page 63]{hastie:tibshirani:friedman:2009}. We consider the models obtained by AIC, BIC, and best subset selection (BSS) criteria, and consider them as our sub-models. They are listed in Table~\ref{tb:prostate:models}.
\begin{table}[h]
 \caption{Full and candidate sub-models for prostate data.} \label{tb:prostate:models}
\centering
\medskip
\begin{tabular}{lp{4in}} \hline
 Selection  & \\
Criterion & Model: Response \~{} Covariates \\
\hline
Full Model & {\tt lpsa}\~{} {\tt lcavol} + {\tt lweight} + {\tt svi} + {\tt lbph} + {\tt age} + {\tt lcp} + {\tt gleason} + {\tt pgg45} \\
AIC & {\tt lpsa}\~{} {\tt lcavol} + {\tt lweight} + {\tt svi} +  {\tt lbph} + {\tt age} \\
BIC & {\tt lpsa}\~{} {\tt lcavol} + {\tt lweight} + {\tt svi} \\
BSS & {\tt lpsa}\~{} {\tt lcavol} + {\tt lweight} \\
\hline
\end{tabular}
\end{table}

Average prediction errors, and their standard deviations for pretest and  shrinkage estimators for various sub-models are shown in Table~\ref{tb:prostate:est}. Prediction errors are based on five- and ten-fold cross validation. Average and standard errors are obtained after repeating the process 5000 times.
\begin{table}[h]
  \centering
\caption{Average prediction errors for various estimators based on $K$-fold cross validation repeated 5000 times for prostate data. Numbers in smaller font are the corresponding standard errors.}\label{tb:prostate:est}
\medskip
\begin{tabular}{l r@{.}l r@{.}l c r@{.}l r@{.}l} \hline
& \multicolumn{4}{c}{Raw CVE} && \multicolumn{4}{c}{Bias Corrected CVE} \\
\cline{2-5}  \cline{7-10}
Estimator & \multicolumn{2}{c}{$K=5$} & \multicolumn{2}{c}{$K=10$}
&& \multicolumn{2}{c}{$K=5$} & \multicolumn{2}{c}{$K=10$} \\
\hline
UR	& &556$_{.030}$ & &548$_{.018}$ && &543$_{.026}$ & &542$_{.017}$ \\\\
R(AIC)	& &535$_{.023}$ & &529$_{.014}$ && &525$_{.020}$ & &523$_{.013}$ \\
R(BIC)	& &537$_{.020}$ & &533$_{.012}$ && &529$_{.018}$ & &529$_{.011}$ \\
R(BSS)	& &582$_{.017}$ & &578$_{.010}$ && &576$_{.015}$ & &576$_{.009}$ \\\\
PS(AIC)	& &554$_{.029}$ & &547$_{.018}$ && &540$_{.025}$ & &541$_{.017}$ \\
PS(BIC)	& &546$_{.026}$ & &541$_{.016}$ && &533$_{.023}$ & &535$_{.015}$ \\
PS(BSS)	& &549$_{.026}$ & &542$_{.016}$ && &536$_{.023}$ & &536$_{.015}$ \\\\
PT(AIC)	& &536$_{.024}$ & &529$_{.014}$ && &526$_{.021}$ & &525$_{.014}$ \\
PT(BIC)	& &538$_{.021}$ & &533$_{.012}$ && &529$_{.019}$ & &529$_{.011}$ \\
PT(BSS)	& &599$_{.030}$ & &601$_{.024}$ && &602$_{.036}$ & &605$_{.029}$ \\
\hline
\end{tabular}
\end{table}

Looking at the bias corrected cross validation estimate of the prediction errors, on an average, restricted and the pretest estimators based on AIC have the smallest prediction errors. This is followed by pretest and the restricted estimators based on BIC. Interestingly, average prediction errors based on the sub-model given by BSS is much higher than those obtained from the models based on AIC or BIC. For instance, restricted model based on BSS has average prediction error 0.576, and the same for pretest estimator is 0.605. For the same sub-model, positive-shrinkage estimator has average prediction error 0.536, which is much less than R(BSS), and PT(BSS). Clearly, positive shrinkage estimator is beating the restricted and pretest estimators for this sub-model. This is a classic example where utility of positive-shrinkage estimator is practically realized. Restricted and/or pretest estimation may perform better under correct specification of the model (e.g., the models given by AIC and BIC for this data set), whereas, positive-shrinkage estimator is less sensitive to model misspecification.

Apparently, in the presence of imprecise subspace information, restricted and pretest estimators fail to produce the best estimates that reduce average prediction errors. On the other hand, positive-shrinkage estimator maintains a steady risk-superiority under model misspecification. This behaviour is illustrated in more detail through a Monte Carlo study in section \ref{sec:simulation}.

\subsection{State Data}
\cite{faraway:2002} illustrated variable selection methods on a data set called {\tt state}. There are 97 observations (cases) on 9 variables. The variables are: population estimate as of July 1, 1975; per capita income (1974); illiteracy (1970, percent of population); life expectancy in years (1969-71); murder and non-negligent manslaughter rate per 100,000 population (1976); percent high-school graduates (1970); mean number of days with minimum temperature 32 degrees (1931-1960) in capital or large city; and land area in square miles. We consider life expectancy as the response. It was found that population, murder, high school graduates, and temperature produce the best model based on AIC or BIC. A model based on CP statistic that includes population, high school graduates, and temperature showed the largest adjusted $R^2$. All the models are listed in Table~\ref{tb:state:full:sub}.
\begin{table}[h]
  \centering
\caption{Full and candidate sub-models for state data.} \label{tb:state:full:sub}
\medskip
\begin{tabular}{lp{4in}} \hline
Selection  & \\
Criterion & Model: Response \~{} Covariates \\
\hline
Full & {\tt Life.exp}\~{} {\tt Population} +  {\tt Murder} + {\tt Hs.grad} + {\tt Frost} + {\tt Income} + {\tt Illiteracy} + {\tt Area} \\
AIC/BIC & {\tt Life.exp}\~{} {\tt Population} +  {\tt Murder} + {\tt Hs.grad} + {\tt Frost} \\
CP & {\tt Life.exp}\~{}  {\tt Murder} + {\tt Hs.grad} + {\tt Frost} \\
\hline
\end{tabular}
\end{table}

\begin{table}[h]
  \centering
\caption{Average prediction errors (thousands) for various estimators based on $K$-fold cross validation, repeated 5000 times for state data. Numbers in smaller font are the corresponding standard errors.}\label{tb:state:est}
\medskip
\begin{tabular}{l r@{.}l r@{.}l c r@{.}l r@{.}l} \hline
& \multicolumn{4}{c}{Raw CVE} && \multicolumn{4}{c}{Bias Corrected CVE} \\
\cline{2-5}  \cline{7-10}
Estimator & \multicolumn{2}{c}{$K=5$} & \multicolumn{2}{c}{$K=10$}
&& \multicolumn{2}{c}{$K=5$} & \multicolumn{2}{c}{$K=10$} \\
\hline
UR	& &879$_{.144}$ & &847$_{.086}$ && &819$_{.119}$ & &820$_{.079}$ \\\\
R(AIC)	& &637$_{.063}$ & &614$_{.036}$ && &599$_{.052}$ & &597$_{.033}$ \\
R(CP)	& &639$_{.058}$ & &639$_{.033}$ && &626$_{.048}$ & &626$_{.031}$ \\\\
PS(AIC)	& &740$_{.124}$ & &690$_{.074}$ && &696$_{.104}$ & &671$_{.068}$ \\
PS(CP)	& &768$_{.106}$ & &746$_{.063}$ && &727$_{.090}$ & &727$_{.058}$ \\\\
PT(AIC)	& &637$_{.066}$ & &614$_{.036}$ && &599$_{.054}$ & &597$_{.033}$ \\
PT(CP)	& &662$_{.069}$ & &639$_{.035}$ && &629$_{.059}$ & &626$_{.032}$ \\
\hline
\end{tabular}
\end{table}

When the models are correctly specified, it is obvious that restricted estimator will perform the best. Such is the scenario for the state data, where the model given by AIC and BIC are the same, and the restricted estimator has the smallest prediction error. Under model uncertainty, however, the scenario will change completely as restricted estimator becomes unbounded when the sub-model deviates from the true structure. This is explored in the simulation study presented in  section~\ref{sec:simulation}. For the correctly specified models, such as in Table~\ref{tb:state:est}, we see that restricted and pretest estimators have the smallest average prediction errors for both five-fold and ten-fold cross validation. The bias corrected version of the cross validation errors are exactly the same for the restricted and pretest estimators.

\subsection{Galapagos Data}


\cite{faraway:2002} analyzed the data about species diversity on the Galapagos islands. The Galapagos data contains 30 rows and seven variables. Each row represents an island, and the covariates represent various geographic measurements. The relationship between the number of species of tortoise and several geographic variables is of interest. The data set has the following covariates: {\tt Species} represents the number of species of tortoise found on the island, {\tt Endemics} represents the number of endemic species, {\tt Area} represents the area of the island (km$^2$), {\tt Elevation} measures the highest elevation of the island (m), {\tt Nearest} is the distance from the nearest island (km), {\tt Scruz} measures the distance from Santa Cruz island (km), {\tt Adjacent} measures the area of the adjacent island (km$^2$). The original data set contained missing values for some of the covariates, which have been imputed by \cite{faraway:2002} for convenience.

The full model and the sub-models based on AIC and BIC are shown in Table~\ref{tb:galapagos:full:sub}.

\begin{table}[h]
  \centering
\caption{Full and candidate sub-models for Galapagos data.} \label{tb:galapagos:full:sub}
\medskip
\begin{tabular}{lp{4in}} \hline
Selection  & \\
Criterion & Model: Response \~{} Covariates \\
\hline
Full & {\tt Species}\~{} {\tt Endemics} +  {\tt Area} + {\tt Elevation} + {\tt Nearest} + {\tt Scruz} + {\tt Adjacent} \\
AIC & {\tt Species}\~{} {\tt Endemics} +  {\tt Area} + {\tt Elevation} \\
BIC & {\tt Species}\~{} {\tt Endemics} \\
\hline
\end{tabular}
\end{table}

We obtain restricted, pretest, and positive-shrinkage estimates of the regression parameters of the Galapagos data. Average prediction errors along with their standard errors for unrestricted (UR), restricted (R), positive-shrinkage (PS), and preliminary test or pretest (PT) estimators are presented in Table~\ref{tb:galapagos:est}. Prediction errors and the standard errors are shown in thousands. PS(AIC) represents positive shrinkage estimates based on sub-model given by AIC, and PS(BIC) represents the same based on BIC. PT(AIC) and PT(BIC) are similarly defined for pretest estimators.

\begin{table}[h]
  \centering
\caption{Average prediction errors (thousands) for various estimators based on $K$-fold cross validation, repeated 5000 times for Galapagos data. Numbers in smaller font are the corresponding standard errors.}\label{tb:galapagos:est}
\medskip
\begin{tabular}{l r@{.}l r@{.}l c r@{.}l r@{.}l} \hline
& \multicolumn{4}{c}{Raw CVE} && \multicolumn{4}{c}{Bias Corrected CVE} \\
\cline{2-5}  \cline{7-10}
Estimator & \multicolumn{2}{c}{$K=5$} & \multicolumn{2}{c}{$K=10$}
&& \multicolumn{2}{c}{$K=5$} & \multicolumn{2}{c}{$K=10$} \\
\hline
UR	& 13&87$_{8.36}$ & 12&63$_{4.36}$ && 11&31$_{6.70}$ & 11&48$_{3.93}$ \\ \\
R(AIC)	& 12&45$_{6.96}$ & 11&62$_{4.28}$ && 10&10$_{5.57}$ & 10&53$_{3.85}$ \\
R(BIC)	& 1&78$_{0.59}$ & 1&65$_{0.24}$ && 1&46$_{0.43}$ & 1&51$_{0.29}$ \\ \\
PS(AIC)	& 13&19$_{7.82}$ & 11&98$_{4.29}$ && 10&75$_{6.27}$ & 10&88$_{3.87}$ \\
PS(BIC)	& 9&07$_{6.53}$ & 7&96$_{3.75}$ && 7&54$_{5.24}$ & 7&32$_{3.38}$ \\ \\
PT(AIC)	& 12&50$_{6.98}$ & 11&63$_{4.29}$ && 10&14$_{5.58}$ & 10&54$_{3.86}$ \\
PT(BIC)	& 5&39$_{7.56}$ & 3&90$_{6.16}$ && 4&40$_{6.08}$ & 3&55$_{5.56}$ \\
\hline
\end{tabular}
\end{table}

For this example as well, since we have selected our sub-models based on AIC or BIC, they are likely to be true, which results in restricted and pretest estimators being the best estimators in terms of prediction errors. We notice that, models based on BIC are smaller in size, and their average prediction errors are smaller than those of the AIC models. The difference in average prediction errors for the two sub-models is noticeably large. Such a large difference between the competing sub-models shows us about the uncertainty in model specification, and the consequences that it cause. Monte Carlo study conducted later in the paper (section~\ref{sec:simulation}) reveals the sensitivity of restricted and pretest estimators, and shows that pretest and restricted estimators are outperformed by positive-shrinkage estimators when the underlying model is misspecified.

It is noted here that the prediction errors are unusually large for this data set. This indicates that the predictors are not quite capturing the variability in the response.

%

\section{Simulation Studies}\label{sec:simulation}

Monte Carlo simulation experiments have been conducted to examine the
quadratic risk performance of positive-shrinkage and pretest estimators. We simulate the response from the following model:
\[
 y_i=x_{1i}\beta_1+x_{2i}\beta_2+\ldots, + x_{pi}\beta_p +
\varepsilon_i, \ \ i=1,\ldots, n,
\]
where
$x_{1i}=(\zeta^{(1)}_{1i})^2+\zeta^{(1)}_{i}+ \xi_{1i}$,
$x_{2i}=(\zeta^{(1)}_{2i})^2+\zeta^{(1)}_{i}+
2\xi_{2i}$, $x_{si}=(\zeta^{(1)}_{si})^2+\zeta^{(1)}_{i}$
with $\zeta^{(1)}_{si} \ \mbox{i.i.d.}\ \sim N(0,1)$, $\zeta^{(1)}_{i}\
\mbox{i.i.d.}\ \sim N(0,1)$, $\xi_{1i}\sim$Bernoulli(0.45) and $\xi_{2i}\sim$Bernoulli(0.45)
 for all $s=3,\ldots, p$ and $i=1,\ldots, n$.
Moreover, $\varepsilon_i$ are i.i.d. $N(0, 1)$.

We are interested in testing the hypothesis $H_0:
\bm{\beta}_j=\bm{0},$ for $j=p_1+1, p_1+2, \ldots, p_1+p_2, $ with
$p=p_1+p_2.$ Accordingly, we partition the regression coefficients as $\bm{\beta}=(\bm{\beta}_1, \bm{\beta}_2)= (\bm{\beta}_1, \bm{0})$. We show results for $\bm{\beta}_1= (1, 1, 1)$, and $\bm{\beta}_1= (1, 1, 1, 1)$ only.

The number of simulations were initially varied. Finally, each
realization was repeated 2000 times to obtain stable results. For each
realization, we calculated bias of the estimators. We defined
$\Delta = ||\bm{\beta} - \bm{\beta}^{(0)}||,$  where
$\bm{\beta}^{(0)}= (\bm{\beta}_1, \bm{0})$, and $||\cdot||$ is the
Euclidean norm. To determine the behavior of the estimators for
$\Delta >0,$ further data sets were generated from those
distributions under local alternative hypothesis. Various $\Delta$ values between [0,1] have been considered.

The risk performance of an estimator of $\bm{\beta}_1$ was measured by comparing its MSE with that of the unrestricted estimator as defined below:
\begin{equation}\label{eq:rmse}
\textrm{RMSE}(\hbbeta_1^{\textrm{UR}} : \hbbeta_1^\textrm{*})=\frac{\textrm{MSE}(\hbbeta_1^{\textrm{UR}})}{\textrm{MSE}(\hbbeta_1^\textrm{*})},
\end{equation}
where $\hbbeta_1^\textrm{*}$ is one of the estimators considered in this study. The amount by which an RMSE is larger than
unity indicates the degree of superiority of the estimator
$\hat{\bm{\beta}_1^\textrm{*}}$ over $\hbbeta_1^{\textrm{UR}}.$

\begin{figure}[h]
\centering
\includegraphics[angle=270, width=5in]{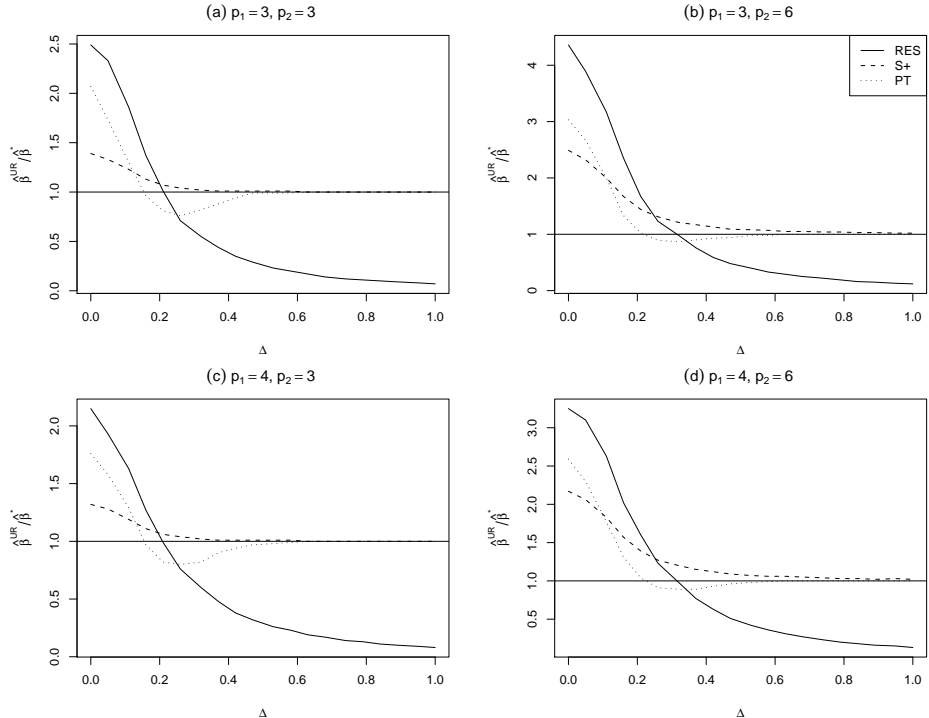}
\caption{Relative mean squared error for restricted, positive-shrinkage, and pretest estimators for $n=50$, and ($p_1, p_2)=$ (3, 3), (3, 6), (4, 3), and (4, 5)} \label{fig:rmse:shrinkage:pretest}
\end{figure}

RMSEs for the positive-shrinkage and pretest estimators were computed for $n=30, 50, 100$, $p_1=3, 6, 9$, and $p_2=4, 6, 9$. Since the results are similar for all the configurations, we list the RMSEs in Table~\ref{tb:shrinkage:pretest:50:4:6} for $n=50$. Comparative RMSEs for positive-shrinkage and pretest estimators for ($p_1, p_2)=$ (3, 3), (3, 6), (4, 3), and (4, 6) are illustrated in Figure~\ref{fig:rmse:shrinkage:pretest}.

\begin{table}[h]
\centering
  \caption{Simulated relative mean squared error for restricted, positive-shrinkage, and pretest estimators with respect to unrestricted estimator
for $p_1=4,$ and $p_2=6$ for different $\Delta$ when $n=50$.}\label{tb:shrinkage:pretest:50:4:6}
\bigskip
\begin{tabular}{cccc}
  \hline 
\rule{0pt}{3ex}  $\Delta^{*}$ &  $\hat{\bm\beta}^{\textrm R}_1$ & $\hat{\bm{\beta}}^{\textrm{S+}}_1$ &  $\hat{\bm{\beta}}^{\textrm{PT}}$  \smallskip
\\ \hline
 0.00 & 3.25 &   2.17 &  2.59 \\
 0.05 & 3.10 &   2.06 &  2.30 \\
 0.11 & 2.63 &   1.83 &  1.77 \\
 0.16 & 2.02 &   1.57 &  1.31 \\
 0.21 & 1.60 &   1.39 &  1.04 \\
 0.26 & 1.23 &   1.27 &  0.91 \\
 0.32 & 0.98 &   1.20 &  0.89 \\
 0.37 & 0.77 &   1.15 &  0.89 \\
 0.42 & 0.63 &   1.12 &  0.93 \\
 0.47 & 0.51 &   1.09 &  0.96 \\
 0.53 & 0.42 &   1.07 &  0.98 \\
 0.58 & 0.36 &   1.06 &  0.99 \\
 0.63 & 0.31 &   1.06 &  1.00 \\
 0.68 & 0.27 &   1.05 &  1.00 \\
 0.74 & 0.23 &   1.04 &  1.00 \\
 0.79 & 0.20 &   1.03 &  1.00 \\
 0.84 & 0.18 &   1.03 &  1.00 \\
 0.89 & 0.16 &   1.02 &  1.00 \\
 0.95 & 0.15 &   1.03 &  1.00 \\
 1.00 & 0.13 &   1.02 &  1.00 \\
\hline
\end{tabular}
\end{table}

\subsection{Case 1: $\Delta=0$}
Clearly, for $\Delta=0$, the restricted estimator outperforms all other estimators for all the cases considered in the simulation study. As the restriction moves away from $\Delta=0,$ the restricted estimator becomes unbounded (see the sharply decaying curve that goes below the horizontal line at $\hbbeta_1^{\textrm{UR}}/\hbbeta_1^{\textrm{*}}$=1 for $\Delta > 0$). The positive-shrinkage estimator approaches 1 at the slowest rate (for a range of $\Delta$) as we move away from $\Delta=0$. This indicates that in the event of imprecise subspace information (i.e., even if $\bbeta_2 \ne  \bm{0}$), it has the smallest quadratic risk among all other estimators for a range of $\Delta$. Pretest estimator outshines shrinkage estimators when $\Delta$ is in the neighbourhood of zero. Otherwise, it becomes unbounded at a faster rate than the restricted estimator. However, with the increase of $\Delta$, at some point, RMSE of pretest estimator approaches 1 from below. This phenomenon suggests that neither pretest nor restricted estimator is uniformly better than the other when $\Delta>0$.

\subsection{Case 2: $\Delta>0$}

Simulation results suggest that positive shrinkage estimator maintains its superiority over the restricted and pretest estimators for a wide range of $\Delta$. In particular, when $p_2=3$, the performance of positive-shrinkage estimator is superior for $\Delta$ up to around 0.35, after which point it is as good as the unrestricted estimator (panels a) and c) in Figure~\ref{fig:rmse:shrinkage:pretest}). However, when $p_2=6$, positive-shrinkage estimator maintains its risk-superiority over all other estimators for a wider range of $\Delta$ (see panels b) and d) in Figure~\ref{fig:rmse:shrinkage:pretest}). This clearly suggests that a positive-shrinkage estimator is preferred as there always remains uncertainty in specifying statistical models correctly. Moreover, one cannot go wrong with the positive-shrinkage estimators even if the assumed model is grossly wrong. In such cases, the estimates are as good or equal to the unrestricted (i.e., full model) estimates.

In the following sections, we review the asymptotic properties of the estimators, and analytically present their bias and risk expressions.

\section{Asymptotic Distribution of the Estimators}

In this section we present the asymptotic distributions of the estimators, and the test statistic $\psi_n$. This facilitates in finding the asymptotic distributional bias (ADB), asymptotic quadratic distributional bias (AQDB), and quadratic risk (AQDR) of the estimator of $\bbeta$.

Under fixed alternative, the asymptotic distribution of $\sqrt{n}(\bbeta^{*} - \bbeta)/s_e$ is equivalent to $\sqrt{n}(\hbbetaUR - \bbeta)/s_e$. This suggest that in asymptotic setup, there is not much to investigate under a fixed alternative such as $\H\bbeta \ne \h$. Therefore, to obtain meaningful asymptotics, a class of local alternatives, $\{K_n\}$, is considered, which is given by
\begin{equation}\label{eq:local:alternative}
K_n: \H\bbeta=\h + \frac{\bm{\omega}}{\sqrt{n}},
\end{equation}
where $\bm{\omega}= (\omega_1, \omega_2, \cdots, \omega_{p_2})'  \in \mathfrak{R}^{p_2}$ is a fixed vector. We notice that $\bm\omega=\bm 0$ implies $\H\bbeta = \h,$ i.e., the fixed alternative is a particular case of (\ref{eq:local:alternative}). In the following, we evaluate the performance of each estimators under local alternative.

 For an estimator $\bbeta^{*}$ and a positive-definite matrix $\bm{W}$, we define the loss function of the form
\[
 L(\bbeta^{*}; \bbeta) = n(\bbeta^{*}- \bbeta)'\bm{W}(\bbeta^{*}- \bbeta).
\]
These loss functions are generally known as weighted quadratic loss functions, where $\bm W$ is the weighting matrix. For $\bm W = \bm I$, it is the simple squared error loss function.

The expectation of the loss function
\[
 E[L(\bbeta^{*}, \bbeta); \bm W] = R[(\bbeta^{*}, \bbeta); \bm W],
\]
is called the risk function, which can be written as
\begin{align}
 \nonumber R(\bbeta^{*}, \bbeta); \bm W)& = n E[(\bbeta^{*}-\bbeta)' \bm W (\bbeta^{*} - \bbeta)] \\
\nonumber & = n \, \textrm{tr} [\bm W\{E(\bbeta^{*}-\bbeta)(\bbeta^{*}-\bbeta)' \}] \\
& = \textrm{tr} (\bm W \bm \Gamma^{*}),
\end{align}
where $\bm \Gamma^{*}$ is the covariance matrix of $\bbeta^{*}$.

The performance of the estimators can be evaluated by comparing the risk functions with a suitable matrix $\bm W$. An estimator with a smaller risk is preferred. The estimator $\bbeta^{*}$ will be called inadmissible if there exists another estimator $\bbeta^{0}$ such that
\begin{equation}\label{eq:inadmissible:estimator}
 R(\bbeta^{0}, \bbeta) \leq R(\bbeta^{*}, \bbeta) \quad \forall (\bbeta, \bm W)
\end{equation}
with strict inequality holds for some $\bbeta$. In such case, we say that the estimator $\bbeta^{0}$ dominates $\bbeta^{*}$. If, however, instead of (\ref{eq:inadmissible:estimator}) holding for every $n$, we have
\begin{equation}\label{eq:asymptotic:idadmissibility}
 \lim_{n \ra \infty} R(\bbeta^{0}, \bbeta) \leq \lim_{n \ra \infty} R(\bbeta^{*}, \bbeta) \quad \forall \bbeta,
\end{equation}
with strict inequality for some $\bbeta$, then $\bbeta^{*}$ is termed as asymptotically inadmissible estimator of $\bbeta$. The expression in (\ref{eq:inadmissible:estimator}) is not easy to prove. An alternative is to consider the asymptotic distributional quadratic risk (ADQR) for the sequence of local alternative $\{ K_n \}$.

Consider the asymptotic cumulative distribution function (cdf) of $\sqrt{n}(\bbeta^{*} - \bbeta)/s_e$ under $\{ K_n \}$ exists, and defined as
\[
 G(\bm y) = \lim_{n \ra \infty} P[\sqrt{n}(\bbeta^{*} - \bbeta)/s_e \le \bm y].
\]
This is known as the asymptotic distribution function (ADF) of $\bbeta^{*}$. Further let
\[
 \Gamma = \int \int \cdots \int \bm{y} \bm{y}' G(\bm y)
\]
be the dispersion matrix which is obtained from ADF, the ADQR may be defined as
\begin{equation}\label{eq:adqr}
 R(\bbeta^{*}; \bbeta) = \textrm{tr}(\bm W \bm \Gamma).
\end{equation}

An estimator $\bbeta^{*}$ is said to dominate an estimator $\bbeta^{0}$ asymptotically if $R(\bbeta^{*}; \bbeta) \le R(\bbeta^{0}; \bbeta)$. Further, $\bbeta^{*}$ strictly dominates $\bbeta^{0}$ if $R(\bbeta^{*}; \bbeta) < R(\bbeta^{0}; \bbeta)$ for some $(\bbeta, \bm W)$. The asymptotic risk may be obtained by replacing $\bm \Gamma$ with the limit of the actual dispersion matrix of $\sqrt{n}(\bbeta^{*}-\bbeta)$ in the ADQR function. However, this may require some extra regularity conditions. \citet{sen:1986}, and \citet{saleh:sen:1985} among others, have explained this point in various other contexts.

%
%
%
%
%
\subsection{Asymptotic Bias and Risk Performance}\label{sec:bias:risk}

To obtain the asymptotic distribution of the proposed estimators, and the test statistic $\psi_n$, we consider the following theorem.
\begin{theorem}\label{thm:saleh:eq:7.8.9}
 Under the regularity conditions, and if $\sigma^2< \infty$, as $n\ra \infty$,
\[
 \sqrt{n} \ s_e^{-1}(\hbbetaUR-\bbeta) \stackrel{d}{\sim}N_p(\bm 0, \C^{-1}).
\]
\end{theorem}
\subsubsection{Bias Performance}
The asymptotic distributional bias (ADB) of an estimator $\bm{\delta}$ is defined as
$$\mbox{ADB}(\bm{\delta})=\lim_{n\to\infty}E\left\{n^{\frac 1
2}(\bm{\delta}-\bm{\beta}_1)\right\}.$$

\begin{theorem} \label{thm:4.1} Under the assumed regularity conditions and theorem above, and under  $\{K_n\}$, the ADB of the estimators are as
follows:
\begin{align}
 \mbox{ADB}(\hbbeta^{\textrm{UR}}_1)&={\bm 0}\\
 \mbox{ADB}(\hbbeta^{\textrm R}_1)&=-\C_n^{-1}\H\bm{B}^{-1}{\bm\omega}\\
\mbox{ADB}(\hbbeta^{\textrm{PT}}_1) &= -\C^{-1}\H\bm{B}^{-1}\delta\H_{p_2+2}(\chi^2_{p_2, \alpha}; \Delta)\\
\nonumber \mbox{ADB}(\hbbeta^{\textrm{S+}}_1) &= -\C^{-1}\H\bm{B}^{-1}\bm\omega\left[\H_{p_2+2}(p_2-2; \Delta) + (p_2-2)E\left\{\chi^{-2}_{p_2+2}(\Delta) \right\}\right. \\
& \left. + E\left\{\chi^{-2}_{p_2+2}(\Delta)I(\chi^2_{p_2+2}(\Delta) > p_2-2) \right\}\right]
\end{align}
\end{theorem}
where
$$E(\chi^{-2j}_p(\Delta))=\int^{\infty}_0 x^{-2j}d\Phi_p(x; \Delta)$$ and $\Phi_p(x; \Delta)$ is the cdf of a $p$-variate normal distribution with mean vector $\bm{0}$, and covariance matrix, $\Delta$.

The bias expressions for all the estimators are not in the scalar form. We
therefore take recourse by converting  them into the quadratic form. Let us
define the asymptotic quadratic distributional bias (AQDB) of an estimator
$\bm{\delta}$ of $\bm{\beta}_1$ by
\begin{align*}
 AQDB(\bm \delta) & = [ADB(\bm \delta)]'\bm{\Sigma} [ADB(\bm \delta)] \quad
\end{align*}
where $\bm{\Sigma}^{-1} = \sigma^2 \C^{-1}$ is the dispersion matrix of $\hbbetaUR$ as $ n \ra \infty$.

Using the definition, and following \cite{ahmed:1997}, the asymptotic quadratic distributional bias of the various estimators are presented below.
\begin{align}
\mbox{AQDB}(\hbbetaUR_1)&={\bm 0},\\
\mbox{AQDB}(\hbbetaR_1) & = \frac{\bm \xi'\bm \xi}{\sigma^2 \C^{-1}} = \Delta\\
\mbox{AQDB}(\hbbeta^{\textrm{PT}}_1) &=\Delta \left\{ \H_{p_2+2}(\chi^2_{p_2, \alpha}; \Delta)\right\}^2\\
\nonumber ADQB(\hbbetaS+_1) & = \Delta\left[\H_{p_2+2}(p_2-2; \Delta) + (p_2-2) E\left\{\chi^{-2}_{p_2+2}(\Delta)\right\}\right. \nonumber \\
& \left. + E\left\{\chi^{-2}_{p_2+2}(\Delta)I(\chi^2_{p_2+2}(\Delta) > p_2-2) \right\}\right].
\end{align}

\subsubsection{Risk Performance}\label{sec:risk:performance}

Following \cite{ahmed:1997}, we present the risk expressions of the estimators.
\begin{theorem}
 Under the assumed regularity conditions, and local alternative $\{K_n\}$, the ADQR expressions are as follows:
\end{theorem}
\begin{align}
 R(\hbbetaUR_1; \bm{W}) & = \sigma^2 \textrm{tr}(\bm{W}\C^{-1}) \\
R(\hbbetaR_1; \bm{W}) & = \sigma^2 \textrm{tr}(\bm{W}\C^{-1}) - \sigma^2\textrm{tr}(\bm{Q}) + \bm \omega'\bm{B^{-1}}\bm{Q}\bm{\omega}\\
\nonumber R(\hbbetaSh_1; \bm{W}) & = \sigma^2\textrm{tr}(\bm W \C^{-1}) - (p_2-2)\sigma^2\textrm{tr}(\bm {Q}_{11})\left\{ 2E[\chi^{-4}_{p_2+4}(\Delta)] \right.\\
& \left. \quad - (p_2-2)E[\chi^{-4}_{p_2+4}(\Delta)]\right\} + (p_2-2)(p_2+6)(\gamma_1'\bm{Q}_{11}\gamma_1)E[\chi^{-4}_{p_2+4}(\Delta)]\\
\nonumber R(\hbbeta^{\textrm{PT}}_1; \bm{W}) & = \sigma^2\textrm{tr}(\bm W\C^{-1}) - \sigma^2\textrm{tr}(\bm Q)\H_{p_2+2}(\chi^2_{p_2, \alpha}; \Delta)  \\
& + \bm {\omega}'\bm {B}^{-1}\bm{\omega}\left\{2 \H_{p_2+2}(\chi^2_{p_2, \alpha}; \Delta) - \H_{p_2+4}(\chi^2_{p_2, \alpha}; \Delta) \right\} \\
\nonumber R(\hbbeta^{\textrm{S+}}_1; \bm{W}) & = \nonumber R(\hbbeta^{\textrm{S}}_1; \bm{W}) + (p_2-2)\sigma^2\textrm{tr}(\bm Q)\left[ E\left\{\chi^{-2}_{p_2+2}(\Delta)I(\chi^2_{p_2+2}(\Delta)\leq p_2-2)\right\}\right. \\
\nonumber & - \left. (p_2-2)E\left\{\chi^{-4}_{p+2+2}(\Delta)I(\chi^2_{p_2+2}(\Delta) \leq p_2-2)\right\}\right] \\
\nonumber & - \sigma^2\textrm{tr}(\bm Q)\H_{p_2+2}(p_2-2; \Delta) + \bm{\omega}'\bm{B}^{-1}\bm{Q}\bm{\omega} \left\{2\H_{p_2+4}(p_2-2;\Delta)\right\}\\
\nonumber & - (p_2-2)\bm{\omega}'\bm{B}^{-1}\bm{Q}\bm{\omega}\left[2 E\left\{\chi^{-2}_{p_2+2}(\Delta)I(\chi^2_{p_2+2}(\Delta) \leq
 p_2-2)\right\} \right. \\
\nonumber& \left. -2E\left\{\chi^{-2}_{p_2+4}(\Delta)I(\chi^2_{p_2+4}(\Delta) \leq p_2-2)\right\} \right. \\
& + \left.(p_2-2)E\left\{\chi^{-4}_{p_2+4}(\Delta)I(\chi^{-4}_{p_2+4}(\Delta) \leq p_2-2)\right\} \right],
\end{align}
where $\bm{Q} = \H\C^{-1}\bm{W}\C^{-1}\H'\bm{B}^{-1}$.

\cite{ahmed:1997} have studied the statistical properties of  various shrinkage and pretest estimators. It was remarked that none of the unrestricted, restricted, and pretest estimators is inadmissible with respect to any of the others. However, at $\Delta=0$,
\[
 \hbbeta^{\textrm{R}}_1 \succ \hbbeta^{\textrm{P}}_1 \succ \hbbeta^{\textrm{UR}}_1.
\]
Therefore, for all $(\Delta;\bm W)$ and $p_2 \ge 3$,
\[
 R(\hbbetaS+_1; \bm W) \le R(\hbbetaSh_1; \bm W) \le R(\hbbetaUR_1; \bm W)
\]
is satisfied.
Thus, we conclude that $\hat{\bm{\beta}}^{\textrm S+}_1$
performs better than $\hbbeta_1^{\textrm{UR}}$ in the entire parameter space induced by $\Delta$. The gain in risk over $\hbbeta_1^{\textrm{UR}}$ is substantial when $\Delta=0$ or near.

\section{Discussion}\label{sec:conclusion}

In this paper, we reviewed positive-shrinkage and pretest estimation in the context of a multiple linear regression model. In our study, we presented asymptotic bias and the risk expressions for the estimators.

When we have prior information about certain covariates, shrinkage estimators are directly obtained by combining the full and sub-model estimates. On the other hand, if \emph{a priori} information is not available, shrinkage estimation takes a two-step approach in obtaining the estimates. In the first step, a set of covariates are selected based on a suitable model selection criterion such as AIC, BIC or best subset selection. Consequently, the remaining covariates become nuisance, which forms a parametric restriction on the full model. In the second step, full and sub-model estimates are combined in a way that minimizes the quadratic risk.

To illustrate the methods, three different data sets have been considered to obtain restricted, positive shrinkage, and pretest estimators. Average prediction errors based on repeated cross validation estimate of the error rates shows that pretest and restricted estimators have superior risk performance compared to the unrestricted, and positive-shrinkage estimators when the underlying model is correctly specified. This is not unusual since the restricted estimator dominates all other estimators when the prior information is correct. Since the data considered in this study have been interactively analyzed using various model selection criteria, it is expected that the sub-models consist of the best subsets of the available covariates for the respective data sets. Theoretically, this is equivalent to the case where $\Delta=0$, or very close to zero. The real data examples, however, do not tell us how sensitive are the  prediction errors under model misspecification. Therefore, we conduct Monte Carlo simulation to study such characteristics for positive-shrinkage and pretest estimators under varying $\Delta$, and different sizes of the nuisance subsets.

In Monte Carlo study, we numerically computed relative mean squared errors for the restricted, positive-shrinkage, and pretest estimators with respect to the unrestricted estimator. Our study re-established the fact that the restricted estimator outperforms the unrestricted estimator at or near the pivot ($\Delta=0$). However, as we deviate from the pivot ($\Delta>0$), risk of the  restricted estimator becomes unbounded. Pretest estimator becomes unbounded even faster than the restricted estimator for the cases considered in the simulation. However, as the $\Delta$ increases, pretest estimator performs better for some $\Delta$, and approaches from below to merge with the line where RMSE is unity. On the other hand, positive-shrinkage estimator decays at the slowest rate with the increase of $\Delta$, and perform steadily throughout a wider range of the alternative parameter subspace. In particular, when the nuisance subset is large, positive-shrinkage estimators outperforms all other estimators, which can be seen in panels b) and d) in Figure~\ref{fig:rmse:shrinkage:pretest}.

\subsection{Future directions}

Pretest estimator either selects restricted or unrestricted estimator depending on the significance based on a test statistic, while positive-shrinkage estimator shrinks the covariates towards the restricted subspace. The nuisance subset is ideally a null space when they do not contribute anything towards the estimation process. In this sense, shrinkage estimators resemble penalized estimators such as the least absolute penalty and selection operator, \emph{lasso}. Proposed by \cite{tibshirani:1996}, lasso is a member of the penalized least squares (PLS) family, which
performs variable selection and parameter estimation simultaneously. Lasso estimates are obtained via cyclical coordinate descent algorithm.

Shrinkage estimation does variable selection by shrinking the coefficients towards the restricted sub-space. In doing so, some of the coefficients shrink towards zero, while some over-shrinks--producing a negative sign for the coefficient. The change of sign may be uncomfortable for practitioners, although it does not affect the risk performance. The positive-part shrinkage estimator takes care of the negative part by setting the coefficient to exactly zero. In the process, most of the coefficients are shrunk while some of them are eliminated by shrinking to zero.

Since the introduction of lasso, there has been a tremendous amount of development in lasso and related absolute penalty estimation (APE) during the past one and a half decade. Although the lasso and shrinkage methods have been around for quite some time, little work has been done to compare their relative performance. Recently, \cite{ahmed:doksum:hossain:you:2007} compared positive shrinkage and lasso in a partially linear regression setup. However, no comparative study for shrinkage and absolute penalty estimators in multiple linear regression model has been found in the reviewed literature. We are currently working on this front, and the findings will be disseminated through future communications.

\bibliographystyle{apa}
\bibliography{phdref}
\addcontentsline{toc}{chapter}{References}
\renewcommand{\leftmark}{References} 

 \end{document}